\documentclass[pdflatex,sn-mathphys]{sn-jnl}

\usepackage[printwatermark]{xwatermark}
\usepackage{xcolor}
\usepackage{graphicx}
\usepackage{lipsum}

\usepackage{xcolor}

\usepackage{comment}

\jyear{2021}%

\theoremstyle{thmstyleone}%
\theoremstyle{thmstyletwo}%

\theoremstyle{thmstylethree}%

\usepackage{lipsum}

\begin{document}

\title[Needs and Artificial Intelligence]{\textit{Needs} and Artificial Intelligence}


\author*[1,2]{\fnm{Soheil} \sur{Human}}\email{soheil.human@wu.ac.at}
\author[3]{\fnm{Ryan} \sur{Watkins}}\email{rwatkins@gwu.edu}

\affil[1]{\orgdiv{Sustainable Computing Lab, Institute for Information Systems and New Media}, \orgname{Vienna University of Economics and Business}, \orgaddress{\street{Welthandelsplatz 1}, \city{Vienna}, \postcode{A-1020}, \country{Austria, EU}}}

\affil[2]{\orgdiv{Department of Philosophy}, \orgname{University of Vienna}, \orgaddress{\street{Universitätsstraße 7}, \city{Vienna}, \postcode{A-1100}, \country{Austria, EU}}}

\affil[3]{\orgname{George Washington University}, \orgaddress{\street{G Street NW}, \city{Washington}, \postcode{20052}, \state{DC}, \country{USA}}}

\title[Needs and Artificial Intelligence]{\textit{Needs} and Artificial Intelligence}


\abstract{
\renewcommand{\thefootnote}{\alph{footnote}}
Throughout their history, homo sapiens have used technologies to better satisfy their \textit{needs}. 
The relation between \textit{needs} and \textit{technology} is so fundamental that the US National Research Council defined the distinguishing characteristic of technology as its goal ``to make modifications in the world to meet human needs'' \cite{nationalresearchcouncilu.s.NationalScienceEducation1996}. 
Artificial intelligence (AI) is one of the most promising emerging technologies of our time. Similar to other technologies, AI is expected ``to meet [human] needs''. In this article, we reflect on the relationship between \textit{needs} and AI, and call for the realisation of \textit{needs-aware} AI systems. 
We argue that re-thinking \textit{needs} \textit{for}, \textit{through}, and \textit{by} \textit{AI} can be a very useful means towards the development of realistic approaches for Sustainable, Human-centric, Accountable,  Lawful, and Ethical (HALE) AI systems. 
We discuss some of the most critical gaps, barriers, enablers, and drivers of co-creating future AI-based socio-technical systems in which [human] needs are well considered and met. 
Finally, we provide an overview of potential threats and HALE considerations that should be carefully taken into account, and call for joint, immediate, and interdisciplinary efforts and collaborations.
}




\maketitle

\section{Introduction}
Technology has historically been created to serve the \textit{needs} and desires of humans, and almost exclusively only to serve a \textit{need} or desire of a specific human or group of humans first. 
Artificial Intelligence (AI)---including forms of Artificial Narrow Intelligence (ANI)\footnote{In this article, we focus more on ANI rather than Artificial General Intelligence or Super Intelligence, though many of the considerations could likely apply to these as well.} designed for tasks that are informed by (or seek to inform) human decisions---is increasingly however gaining the capacity to serve much broader ambitions.
In discussing his popular book, \textit{AI 2041: Ten Visions for Our Future}, author Kai-Fu Lee posits that in the future ``AI will learn to serve human needs'' \cite{lee_oreilly_2021}.
Similarly, Human-Centered AI (HCAI) has ``serve human needs'' as a primary application goal \cite{Shneiderman2020HCAIgoals}, and it has been suggested that the defining characteristic of all technologies is their capacity to serve human \textit{needs} \cite{nationalresearchcouncilu.s.NationalScienceEducation1996}.
On the surface, serving human needs appears to be a laudable goal for AI and AI developers, and within reach given the current fast-paced evolution of AI related technologies.
Yet, in this article, we weigh the ethical and pragmatic implications of this ambition--and consider what it would take to make \textit{needs-aware} AI a reality.
After all, currently we do not even have broad agreement(s) across communities, disciplines or cultures on a single definition (or a set of co-existing definitions) of what \textit{needs} are (and are not)\cite{altschuld2014primer, humanOntologyRepresentingHuman2017a}, let alone what constitutes high-priority \textit{needs} for individuals, organizations, or societies.
Nor do we know how AI can assist in determining what responses are going to best \textit{satisfy needs}, or even how \textit{needs satisfaction} is best measured.
From the barriers and technical challenges, to the driving forces that we believe can push societies toward \textit{needs-serving} technological futures, in this first article (of what we hope will be a series of articles by many contributors with diverse perspectives), we start to reflect on (and then co-create) a future where AI systems have the capacity to help us meet \textit{needs}.

There is an urgency to beginning this journey, a ``burning platform'' \cite{nickols2010four} of sorts: more and more AI applications are in development, AI is increasingly important in many aspects of peoples' lives, and AI development \textit{won't necessarily wait} for \textit{needs scholars and practitioners} to sit on the fences of the issues we outline in this article.
AI development is evolving rapidly, and though there is still a great distance to go before artificial general intelligence, today's intelligent agents are already changing lives at home, work, and societies without adequate systematic, comprehensive, or practical ways to integrate the awareness of \textit{needs} into their design, implementation, nor evaluation.
 
Similar to ``intelligence'',  \textit{needs} are difficult to define in a sense that is acceptable for \textit{each and every} one of us\footnote{Sternberg \cite{sternberg2000concept} beautifully formulates this: ``[l]ooked at in one way, everyone knows what intelligence is; looked at in another way, no one does. Put another way, people all have conceptions - which also are called folk theories or implicit theories - of intelligence, but no one knows for certain what it actually is.'' We can replace ``intelligence'' with ``need'' in these sentences and they will still be valid. Defining ``need'' is not easier than defining ``intelligence'', if not harder.}, especially among scholars from different disciplines and schools of thought. Hence, providing an ultimate \textit{one-line} definition of \textit{need} does not seem to be feasible, and may not even be desirable. 
This is, among others, due to 1) the general difficulty of defining ``concepts'' using natural languages (as widely discussed in philosophy and cognitive sciences, see e.g. \cite{laurence1999concepts}), 2) the wide usage of \textit{need} in both common and professional contexts, and 3) the potential complexity and multidimensionality of \textit{needs} and the \textit{knowledge of needs} (see e.g. \cite{mcleod2011knowledge}). And yet, if people can't agree on what \textit{needs} are, and are not, then how can AI systems be expected to serve \textit{needs}.

\textit{Need(s)} in this context is a specific term, just as are the terms ``intelligence'' and ``artificial intelligence''.
The word \textit{need} (especially when used as a noun) is deliberately selected by authors (including us) because it has the connotation of meaning \textit{a[n intrinsic] necessity} for [the well-being or well-functioning of] a system (e.g., a human, a living agent, an organization, a society, etc.).\footnote{Throughout the article we italicize the word ''need'' as a reminder that we are using it to refer to a specific construct, and intentionally not using the word in persuasive manner to imply a lack of alternatives or options.}
This perspective, we hope, can be helpful to distinguish \textit{needs} from other terms such as ``wants'', ``cravings'', ``wishes'', ``motivators'', or ``desires'' in most cases.
We also distinguish \textit{needs} and \textit{satisfiers}.
For example, an individual may have a \textit{need} (e.g., improved nutrition in order to maintain well-functioning) that can be \textit{satisfied} by a specific food (e.g., cauliflower, or carrots) in a specific context (location, time, situation, etc.). 
Here, the person's \textit{need} (more specifically, the difference between the nutrition necessary to maintain their well-functioning and their nutrition level at the time) is not the same as the potential \textit{satisfiers} of that \textit{need} in the described context.
Clearly, the same or different people in different contexts can satisfy a similar \textit{need} (e.g. nutrition) through different satisfiers (e.g. bread, pizza, rice, etc.), and in the future both the \textit{need} and potential satisfiers may very well change a little or quite substantially.
Moreover, a \textit{satisfier} may not always be an object (such as food), but could also be actions and activities (such as ``meeting friends'' or ``exercising''), or a combination of objects and activities.
Additionally, the mapping between \textit{needs} and \textit{satisfiers} (depending on our level of abstraction) can be complex: multiple \textit{needs} can be satisfied by a single or multiple satisfiers, and multiple satisfiers can satisfy single or multiple \textit{needs}. Here--considering these complexities--we find both ethical challenges and contexts in which \textit{needs-aware} AI technologies could be potentially quite helpful.

In the development of the technologies that power AI (and those that are powered by AI), we contend AI-driven socio-technical systems are ideally sustainable. 
Here, we will apply \textit{H}uman-centric, \textit{A}ccountable, \textit{L}awful, and \textit{E}thical AI (Sustainable HALE AI)\footnote{The term \textit{HALE} AI is borrowed from \textit{the HALE WHALE}, recently proposed for the co-creation of Sustainable, Human-centric, Accountable, Ethical, and Lawful socio-technical systems \cite{humanHALEWHALEFramework2022}} as a framework for sustainable AI. 
Noting however that, even here, \textit{needs} should find a more applied role--which is one of our motivations for writing this article.

It has been suggested that AI developers are often placed in \textit{social dilemmas} with societal good on one side and commercial pressures on the other \cite{struemke2021social}.
We submit that part of the solution to resolving these dilemmas, beyond ethical and legal/regulatory  frameworks, is the introduction of measurable \textit{needs} (or measurable \textit{needs satisfaction}) into on-going efforts to achieve AI that is both aware of and helps resolves \textit{needs}.
By identifying and measuring \textit{needs} (e.g., societal, organizational, and individual \textit{needs}), we have the best chance of finding an appropriate equilibrium that serves them meaningfully and in a balanced manner. 
We cannot, however, achieve this by ignoring \textit{needs}, or incorporating them just superfluously without working definitions or standards for what they are, what they are not, how they relate, and how they can/should be enacted, utilized, satisfied or measured.

It is worth mentioning that by \textit{measuring needs} (or measuring needs satisfaction), we do not necessarily mean converting \textit{needs} to \textit{numbers}. \textit{Needs} are implicit constructs, by \textit{measuring} them here, we mean \textit{explicitizing}, \textit{utilizing} or \textit{enactizing} the knowledge of/about \textit{needs}\footnote{or knowledge of/about needs satisfaction}\cite{mcleod2011knowledge} by applying qualitative, quantitative, and mixed methodologies. When it is about measuring needs for/through/by AI, we mean to \textit{make the knowledge of/about needs accessible, useable, enactable} for/through/by AI systems. We think that the recent and emerging advancements in development of AI and other digital technologies can make it possible to apply novel methodologies in this regard.

Whereas some of the existing approaches (e.g. various regression and Bayesian tools especially) have been very successful in creating valuable machine learning tools (such as classifiers) in domain-specific applications, we posit that the future development of Sustainable HALE AI requires additional concepts and tools associated with \textit{needs}.
\textit{Needs} (philosophically, socio-technically, and computationally) is a construct that has the capacity to guide human and societal decisions in creating AI systems, along with guiding machine decisions and behaviours during implementation.
This capacity of \textit{needs} can be applied (along with predictive tools and ethical frameworks) at various phases of AI development and application to create AI that is capable of serving human \textit{needs}.

\textit{Needs} can be \textit{for} AI, \textit{through} AI, and \textit{by} AI.
That is to say, \textit{for AI} refers to understanding \textit{needs} to be used in AI systems, \textit{through AI} refers to understanding \textit{needs} through the process of co-constructing needs-aware AI systems; and \textit{by AI} refers to using AI systems to understand how humans satisfy \textit{needs} (e.g., needs-mining\footnote{Needs-mining analyzes data from social media and other sources in order to identify users \textit{needs}, desires, and preferences.}, mapping \textit{needs} to satisfiers, evaluating \textit{needs} satisfaction).

A primary goal of this article is to initiate a multi-disciplinary and interdisciplinary professional dialogue about what are the appropriate roles for \textit{needs} in the design, development, and application of AI technologies in the coming decades.
We do not propose answers, nor are we naive enough to believe that this can be done overnight.
Rather, we want to focus attention on the valuable role that a measurable (\( \approx\) explicitizable, accessible, utilizable, or enactable) construct of \textit{needs} can have from the design decisions that going into creating a sustainable HALE AI-based socio-technical system, through to the technical weighing of options the systems must do in order to make decisions and/or recommendations.
Moreover, we will reflect on a set of potential challenges, barriers, gaps, drivers, enablers, and considerations regarding the application of \textit{needs} in AI and the development of \textit{HALE needs-aware AI} systems.

\section{The Necessity of (Re-)Introducing \textit{Needs}}

\textit{Needs} have played an essential role throughout the history of philosophy and science. From Aristotle to Marx, many philosophers have used both the concept of \textit{needs}, and the powerful literary tool of the word \textit{need}, as a part of their philosophical frameworks (see \cite{humanOntologyRepresentingHuman2017a} for an overview).
More recently, psychologists, cognitive scientists, social scientists, economists, and experts from many disciplines and sectors have also conceptualized and applied \textit{needs} in practical ways (see  \cite{watkins_2016} for a collection of references).
Similarly, computer scientists and AI experts also continue to consider \textit{needs} in various architectures and systems (e.g., \cite{pearl1999probabilities, humanHowCanPluralist2019,shneiderman2020human}).
While such attempts are precious, in the following, we suggest that now is the time to reinvigorate research and professional dialogues on the roles for \textit{needs} in AI systems (from novel aspects, to multi-dimensional and interdisciplinary approaches, to new measurements). 

\textit{Why now?}
Because we are at a crucial point in the development of ``intelligent'' systems that (when combined with other emerging technologies and approaches) can substantially influence both the  well-being of humans (or even the \textit{being} of humans) and the sustainability of our societies, in the not too distance future. 
We are not fully there yet, so now is the time to solidify \textit{needs} as a measurable construct and input into decisions that can also be used to evaluate our success, so that we (and our machines) can use \textit{needs} in defining and creating a future that we [all] desire.

\textbf{1. The next level co-production}
A vast number of concepts and evidence (from the sociology of science and technology, e.g. \cite{jasanoff2004states, harbers2005inside}, to cognitive science, e.g. \cite{clark1998extended, dror2008offloading}) emphasize that humans and technologies co-produce (i.e., co-create, or co-construct) each other.
AI is no exception--that is, humans and AI systems co-produce each other.
For instance, the dynamic relationships of people and social media recommendation engines (\cite{rowlands2008google, asher2011search}, the shaping of behavior through Internet of Things (IOT; see \cite{elayan2021internet}), the co-evolution of law and technology \cite{eaglin2021critical}, and the changes in how people associate their knowledge in relation to knowledge that always available to them online \cite{ward2021people}).
The emergence of ubiquitous \cite{lyytinen2002ubiquitous, greenfield2010everyware} and pervasive \cite{satyanarayanan2001pervasive} computing has also led to a global web of \textit{ambient intelligence}  \cite{aarts2009ambient, cook2009ambient, gams2019artificial} \textit{everyware} \cite{greenfield2010everyware}, making the importance of this specific human—technology co-production more apparent--and more far-reaching. 
Beyond ubiquity, there are other aspects that make human—AI co-production an important matter of consideration.
For instance, computational systems in general (and AI systems in particular) can embody many capabilities that past technologies could hardly achieve--such as memorizing, computation, inference, decision making, visualization, etc.
With such capabilities, AI can co-create humans'\footnote{as well as groups', communities', organizations', societies', etc.} \textit{needs} and the ways they \textit{satisfy their needs}.
Therefore, we argue, \textit{needs}, and \textit{needs satisfaction}, should be well considered (and studied) in relation to AI development and applications--from initial design decisions and [training] data selection, to development, application, evaluation, and beyond.
But this is just one-side of the \textit{AI--human co-production} of \textit{needs}. The other side is that our understanding of \textit{needs} and the imaginaries (i.e., shared visions and values) we have about them will also contribute to the co-creation of AI systems in the future. 
In other words, AI\footnote{among others} will fuel our dreams of what AI can do, giving us new ideas about what we might want accomplish in the future.

When \textit{needs} are considered, AI and humans do, can, and will  have multiple intersecting relationships.
Humans, for instance, develop AI systems based on perceived or imagined \textit{needs}, they are also routinely the beneficiaries of actions to address \textit{needs}, they identify emerging \textit{needs}, and likewise they are often in the role of assessing current \textit{needs} and evaluating the extent to which \textit{needs} have been satisfied.
For their part, AI systems are just starting to assist people in identifying and prioritizing activities to satisfy \textit{needs}, improving the efficiency of solutions to address \textit{needs}, and at the same time creating new \textit{needs} that didn't exist in prior generations (necessities for both humans and AI systems alike)\footnote{in the context of our increasingly \textit{digital} societies}.
However, AI might play more roles in the coming years, and the weightings of this continuous co-creation might change. 

\textbf{2. AI vs AIs}
There is, of course, no one [human] \textit{need}, as there is no single concept of AI.
Both are complex and contextual, and yet they must consistently interact.  
One current challenge in these relationships is that many recent advancements in AI are mainly based on machine learning approaches, which routinely rely heavily on models and conceptualizations in which  \textit{needs} do not play a central role.
We suggest that AI developers who intend to identify, address, or co-produce \textit{needs} with/for humans (e.g., through HCAI methods) can benefit from integrating \textit{needs} into both their design (such as, identifying which \textit{needs} they intend to address) as well as in their architecture (such as, complementing regression-based ML techniques with necessity and sufficiency analyses).
This is not new to AI either, the AI pioneer Judea Pearl (see e.g., \cite{pearl1999probabilities}) proposed formulas for calculating the probability of necessity and probability of sufficiency; but exploring the role of \textit{needs} has been overshadowed in recent years by \textit{mainstream} approaches. AI is an emerging and evolving field, what is meant by AI and how it is practised, can be different in different domains, contexts, application areas and times. We believe that \textit{re-introducing} \textit{needs} to AI can contribute to develop variants of AI that can better met individual, organizational and societal \textit{needs}.

\textbf{3. Recent interdisciplinary advancements}
While \textit{need} is an old concept and attempts toward considering \textit{needs} in AI systems are likewise not new (see above), recent advancements in disciplines such as cognitive science, sociology of science and technology, and computer science can provide novel concepts, methodologies, and approaches for developing innovative \textit{needs-aware} AI. This however demands, in many cases, a fundamental rethinking about the conceptualizations and implementations of \textit{needs} \textit{for AI}, \textit{through AI}, and \textit{by AI systems}. For example, the recent advancements regarding the \textit{predictive processing} account of cognition \cite{humanEnactiveTheoryNeed2018a} might provide useful concepts and approaches regarding one way of realization of \textit{needs-aware} AI systems--among others. Moreover, in conjunction with these advancements, AI (and AI related technologies) are becoming increasingly commonplace in people's lives. From IoT devices feeding data to ML algorithms that in turn shape people's behaviour \cite{elayan2021internet}, to self-driving cars and AI supported medical decision aids, the expansion of AI into the lives of people requires a renewed focus on how AI systems can co-produce and co-address \textit{needs} of diverse varieties--creating AI that serves human \textit{needs}.  As a consequence, we propose, both \textit{need sciences} (i.e. disciplines that study \textit{needs}) and AI have advanced enough in the last years to construct novel \textit{enabling spaces} \cite{peschl2014designing} for rethinking \textit{need}-AI relations. 

\textbf{4.\textit{Needs} and HALE AI}
While experts from different disciplines and domains have called for the realization of more \textit{human-centric}, \textit{accountable}, \textit{lawful}, and \textit{ethical} (HALE) AI \cite{humanHALEWHALEFramework2022}, how to \textit{realize} such socio-technical digital systems remains very challenging. Since \textit{need} is a fundamental concept that plays an essential role in different aspects of human-centricity, accountability, lawfulness, and ethics, we propose that rethinking \textit{needs} \textit{for}, \textit{through}, and \textit{by} AI can be a very useful means towards the development of \textit{realistic approaches} for HALE AI:

``Both wants and needs are always tied to value prioritizations --  they are not value neutral. Needs evolve within certain historical and cultural contexts.'' \cite{naess2009up} \textit{Needs} can therefore put decisions in historical and cultural contexts, just as those contexts shape what are \textit{needs} at the time. Situational and historical contexts matter immensely as we look to implement \textit{needs-aware HALE AI}, throughout design, development, and implementation. An AI in a hospital during a pandemic will require different consideration of \textit{needs} in comparison at different times, or in comparison to an AI to be employed in managing vending machines, for instance. \textit{Needs} can be informative in both contexts, but contexts would vary how/when \textit{needs} are included into the judgements of both humans and machines.  

Likewise, attempting to develop AI that responds to and/or are responsive to underprivileged communities (whether based on race, gender, ethnicity, economics, or combinations of these and other variables) demands a multi-disciplinary understanding of human \textit{needs}--integrating multiple `levels' (individual, organizational, societal \textit{needs}) \cite{kaufman2019alignment}. 
In other words, \textit{needs} can fundamentally contribute towards shifting the practice of \textit{one-size-fits-all} AI to a more \textit{human-centric}, pluralist, and inclusive approach \cite{humanHowCanPluralist2019}).

\textbf{5. Collective \textit{needs} and digital sustainability}
\textit{Needs} are not limited to individuals.
Teams, organizations, communities, and societies all have \textit{needs} as well (e.g., \cite{kaufman1992strategic, watkins2012guide}).
These collective \textit{needs} are each, and together, relevant to the design and implementation of AI systems.
Identifying \textit{needs} at all levels,  finding solutions to \textit{needs} (plural, in different levels), not sub-optimizing decisions related to one \textit{need} (at one level) at the expense of another \textit{need} (possibly at another level), all have to be considered in order to have digital sustainability.
That is, the \textit{needs} of the individual must be considered in relation to the \textit{needs} of the society, and vice versa across multiple levels.
For example, the use of natural resources is just one example where societal \textit{needs} (for instance, associated with clean energy and climate change) must be consider in relation to the development of AI systems (where training a single large language model can require enormous amounts of electricity, while the outputs of the model may assist many individuals in meeting a personal \textit{need}).

\textbf{6. Diversifying the AI community through \textit{needs}}
The development of commercial AI products is just one avenue for engaging broader audiences in the design and implementation of AI systems (e.g., psychologists and Woebot).
An on-going dialogue on the roles of \textit{needs} in AI is another path that can bring people with diverse perspectives into the AI community; to help guide ethical considerations related to \textit{needs}, to look at policy implications of AI co-production and co-addressing of \textit{needs}, multi-level measurement of \textit{needs} and \textit{needs satisfaction}, and many other essential topics.
From philosophers to social workers, and medical doctors to educators, the topic of \textit{needs} can enlarge and diversify the community of AI researchers, designers, and developers.

\textbf{7. Needs and the AI technoscience}
Today, from mathematics and cognitive science, to economics and cosmology, AI plays a fundamental role by generating data and the means for understanding of many scientific findings.
For instance, in cognitive science the \textit{nature of cognition and interactions} of [cognitive] systems is increasingly explored through AI supported research.
In this sense, AI is not only an engineering approach, but also a partner in many scientific disciplines.
The notion of \textit{AI technoscience} might capture the integration and co-creation of both AI as a technology and AI as a science.
AI as a \textit{technoscience} can highly support the current efforts towards understanding the \textit{nature of [human] needs} and \textit{needs satisfaction} (see e.g. \cite{humanLearningSatisfyNeeds2017,humanEnactiveTheoryNeed2018a} as basic attempts in this direction). For example, AI-based simulations can not only be used as means for the evaluation of the emerging perspectives, but also can inform such perspectives, or even inspire new perspectives towards \textit{needs}.


\section{Gaps and Barriers}
One of the first steps towards rethinking \textit{needs} in AI \footnote{i.e., towards, among others, (i) realization of novel \textit{need-aware AI systems}, (ii) co-construction of new interdisciplinary communities on the intersection of \textit{needs} and \textit{AI}, (iii) co-construction of novel understandings of \textit{needs}} is to identify the existing gaps and barriers that are limiting the use of \textit{needs} \textit{by}, \textit{for}, and \textit{through} AI development today, or making it a difficult challenge.
There are of course many gaps and barriers, we reflect here on a short list with some of the most impactful ones (from our perspective) that should be considered initially--and weighed regularly against others, as well as emerging gaps or barriers.

\textbf{1. Defining \textit{need} and \textit{needs}: a historical challenge}
What is a \textit{need}, what are human \textit{needs}, what are potential categories or classes \textit{needs}, what are the relationships (or potential hierarchies) among \textit{needs}, and how can \textit{needs} be satisfied, have been a topic of inquiry since the time of ancient philosophers (see \cite{humanOntologyRepresentingHuman2017a}).
In the last century, other disciplines diversified the discourse, but no common answers, definition, or agreement within and across different disciplines exist.
From philosophy to economics, and medicine to psychology, \textit{needs} are examined through diverse lens, leading to little agreement on what should (and thereby what should not) be classified as a \textit{need} (or if we can even know our \textit{needs} at all; \cite{mcleod2011knowledge}). 
For example, for economists \textit{needs} are routinely viewed through the lens of income elasticity of demand (see \cite{mccain2014need}), whereas for psychologists the focus is typically on an individual's motivation derived from their \textit{needs} (see \cite{maslow1943theory, ryan2020intrinsic}). 
In health care alone, there are at least five interpretations of what \textit{needs} are \cite{anders2017indeterminacy}.
We do not attempt to seek a solution to this barrier here, rather we simply want to acknowledge that if Kai-Fu Lee proposition that in the future ``AI will learn to serve human needs'' \cite{lee_oreilly_2021} is to become a reality, then we must continuously try to actively reconstruct joint basic understandings or \textit{working definitions} on what \textit{needs} are so that AI can assist in meeting them.
We also do not suggest that a single universal definition is required, or maybe even desired, but rather that the challenge of presenting coherent and valuable [working] definitions for distinct use cases and ways of dealing with diverse (and even disagreeing \cite{humanSupportingPluralismArtificial2018a}) definitions must become a priority for \textit{needs} scholars regardless of their discipline.
The capacity of AI systems to help us meet our \textit{needs} is contingent on our ability to first determine how those systems define and measure \textit{needs}.

\textbf{2. Understanding and applying \textit{need} in the network of concepts}
We also recognize that \textit{need} is not an isolated concept; that is to say, \textit{needs} are usually closely associated with values, rights, desires, wants, preferences, motivations, and other constructs that contribute to shaping our daily perceptions and decisions\footnote{or at least our interpretations or models of such perceptions and decisions}. As a result, it is challenging to model \textit{needs-aware AI systems}--in particular cognitivist ones--\cite{humanHowCanPluralist2019}, without considering other related constructs, which likewise adds to the complexity. Moreover, \textit{needs} are often fluid on multiple dimensions among these concepts as well, which can make the construction of \textit{needs-aware} systems a matter of challenging interpretation and context-based distinction between \textit{needs} and the other constructs. 
This complexity should not, however, dissuade us from the task; rather, we should leverage the capabilities of current technologies to assist us in recognizing and benefiting from the complexity.

\textbf{3. \textit{Needs} vs \textit{satisfiers}}
\textit{Needs} are routinely considered an implicit construct, sometimes informally even defined by ``I know it when I see it'' criteria, although some explicit measurable definitions are also available \cite{watkins2012guide}).
\textit{Satisfiers}, however, are considered explicit; for example, a specific food or liquid that is necessary and sufficient in a specific context (time, location, environment) for a specific person ``in need''.
Based on this perspective, \textit{need satisfaction}--as an action--refers to the \textit{process} of satisfying one or more \textit{needs}. Clearly, \textit{satisfiers} could be both specific objects (e.g., a specific food), specific actions (e.g., talking with a friend in a specific context), or a combination of objects, environments and actions (e.g., a specific party). Though what satisfies a \textit{need} should not be confused, within that relationship, with the \textit{need} itself.
Humans--as embodied cognitive systems that are enacting in their environments--normally \textit{satisfy} a set of \textit{needs} simultaneously while interacting with (i.e., living in) their environments. Understanding and applying the distinction between \textit{needs} (implicit and potential), \textit{satisfiers} (normally, explicit and realized), and \textit{need satisfaction} (normally, explicit realized process, or \textit{`explicitizable'}) can be a challenging task for the co-development of needs-aware AI systems.
If \textit{needs}, \textit{satisfiers}, and \textit{need satisfaction} should be modelled or represented (in an AI system) as \textit{constants}, \textit{variables}, \textit{functions}, \textit{processes}, \textit{decisions}, \textit{actions}, or even as overall \textit{system's dynamics} or \textit{states} (e.g., in line with some of the dynamical models or enactive approaches in cybernetics, system engineering or cognitive science) is a challenging philosophical, scientific, and engineering question.
Moreover, it is important to consider the potential perspectives, or interpretations involved.
For example, by observing someone's behaviours, an AI system (from a third-person perspective) might infer that the person is satisfying a specific \textit{need}, while the person might not have the same opinion or experience (from a first-person perspective).
The same can be valid for AI systems, scientists and experts involved, as well as different individuals or organizations or societies.
To add to this complexity, these relationships of \textit{needs} and satisfiers are continuously co-produced through human--AI interactions (see above).

\textbf{4. Evaluating \textit{needs} and \textit{needs satisfaction}}
Given the conflicting definitions of \textit{needs}--and confusion of terms related to \textit{needs}--an associated barrier to introducing \textit{needs} at various phases of AI development and implementation is limited metrics for measuring\footnote{i.e. \textit{`explicitizing'} \textit{needs} and \textit{needs satisfaction}}, utilizing\footnote{or \textit{`enactizing'} needs} and evaluating \textit{needs} and \textit{needs satisfaction}. Since 1) we are not consistent in defining what is a \textit{need} (and what is not) and 2) the application domains and contexts might vary, we rely on varied measures and utilization mechanisms regarding \textit{needs} and \textit{needs satisfaction}; from very subjective to a fairly objective measurements and mechanisms.
And without measures and utilization mechanisms it might be difficult for AI systems to assist with identifying \textit{needs}, or prioritizing \textit{needs} (within and across multiple levels -- such as, individual, organizational, and societal \textit{needs}), or defining what is required of potential activities to satisfy \textit{needs}, or evaluating when \textit{needs} have been met. Each of these, and others, would represent valuable ways that AI could help serve human \textit{needs}.  


\textbf{5. The dominance of \textit{needs-blind} AI}
Today, we do not expect that the AI systems being developed at Meta (formerly, Facebook), Google, Microsoft, OpenAI, Baidu, or elsewhere will have necessarily any direct knowledge of our \textit{needs}.
We know that some of these technologies might assist us in meeting our \textit{needs}, but this can often be an unintentional secondary benefit--not as a direct result of our \textit{needs} as an `input' into their computations\footnote{We know that the ultimate goal of many companies is to increase their own benefit, and not necessarily satisfying their users' \textit{needs}. As it is famously said, users are in many cases \textit{products} rather than \textit{customers}}.
There is, after all, no technological, legal, or event ethical frameworks or guidelines at this time that could rigorously facilitate AI serving human \textit{needs} if we did suddenly decided to expect such direct benefits (i.e., not just as a potential by-product). 
We agree that HCAI approaches are a positive move toward AI that help meet \textit{needs}, though too often they refer to human \textit{needs} (see, for example, \cite{shneiderman2020human, ahani2021humancentric}) without any definitions, deep reflection, or applicable conceptualization of what those \textit{needs} are, how they would be measured, how they would be satisfied, or how AI developers would know if \textit{needs} were met. Rather HCAI approaches routinely seem to imply the AI developers will consistently and correctly recognize the \textit{needs} of other people (e.g., future users or beneficiaries) through conversation or observation--which is simply inaccurate and unrealistic (as has been the experience of development economics, where a similar approach has been tried for identifying \textit{needs} of those living in the poorest countries of the world). 
Without any tools for systematic assessment of \textit{needs}, at this point in time commercial success dominates what we (as individuals, organizations, and as societies) expect from AI systems.
This can change, but it will require effort; and we propose that re-introducing \textit{needs} into those change efforts is essential.

From a societal perspective, we also recognize that in some (if not most) of the existing socio-economic and power-related structures and institutions, AI development has a tendency to to create wants-serving machinery. We are well aware that, tinkering with the technology alone is not enough to reverse this trend, and AI itself might even conceal structural dynamics, power relations, etc. that reinforce these structures. A potential barrier towards the development of \textit{needs-aware} AI could therefore be \textit{needs-blind} socio-economic perspectives and structures within our institutions. Some of these perspectives or structures are derived from not recognizing the potential of/for \textit{needs-aware} AI; and some of perspective are shaped by a potential conflict of interest with the realization of \textit{needs-aware} AI. We hope that this article, and future contributions by others, can lead toward the development of \textit{imaginaries} and \textit{perspectives} that can enable a shift in our socio-economic perspectives and structures towards \textit{needs-awareness} (in AI and beyond). 

\textbf{6. Missing \textit{needs} in guidelines, standards, regulations, and policies}

Technologies are co-created by different human and non-human actors. Guidelines, standards, regulations, and policies are important non-human actors \cite{lawactor, latour2007reassembling} in the development of any technology, including AI. For many AI-related concepts (e.g., privacy, bias), there are emerging legal requirements, ethical frameworks, or policy mandates, that guide AI developers in making decisions that lead to improved (from a societal perspective) AI systems. We propose that similar efforts to provide valuable guidance to AI developers based on what shared societies want from AI would beneficial for further introduction of \textit{needs}. 

\textit{Need} cannot however just be ``window dressing'' on policies, frameworks, or the like; \textit{needs} must be integrated into the fabric of what we want \textit{for}, \textit{through}, and \textit{by} AI systems. For instance, we cannot just say that AI developers should assess \textit{needs}, but go no further into (i) what \textit{needs} are--and are not, (ii) how \textit{needs} are prioritized, (iii) how \textit{needs} will be assessed, measured, and their satisfaction evaluated, or (iv) how societal \textit{needs} will be balanced with those of individuals. These and other questions must be considered, debated, and revised as we move forward to develop AI systems that have the capacity to help meet our \textit{needs}. Moreover, considering \textit{needs}, and \textit{need-aware AI} systems in the \textit{network of actors and constructs} is critically important to constructing guidelines, standards, regulations, and policies that not only consider \textit{needs} and \textit{needs satisfaction}, but also other related (and interrelated) constructs such as privacy, agency, rights, values, etc. (all together).

\textbf{7. Need-community: A non-existing reality}
From an actor-network perspective, as well as a co-production point of view, communities play an important role in co-realisation of sociotechnical digital systems (including AI systems).
Surrounding \textit{needs} (in engineering, sciences, humanities, social sciences, etc.), however, no global community exists.
The existing small communities also do not have many connections.
Like many topics of research and discussion, \textit{needs} are often debated within the silos of individual disciplines or speciality areas.
Those in philosophy debate \textit{needs} in isolation from those debating \textit{needs} in public health or political science.
Psychology examines \textit{needs} in the context of human motivations whereas the field of human/organizational performance measures \textit{needs} as gaps between current and desired results.
Likewise, while authors like Pearl \cite{pearl1999probabilities} write about \textit{needs} (necessity and sufficiency) in computer science, each of these conversations are disconnected from discussion of \textit{needs} in social work and education.
For AI to assist in meeting \textit{needs}, each of these (and other) communities of scholars and practitioners must come collaborate across silos--which also means collaborating across epistemological divides \cite{cetina1999epistemic} or disagreements \cite{humanSupportingPluralismArtificial2018a}.

These initial gaps we have identified are also not isolated systems, acting on their own.
They are responding to each other, co-creating new gaps, and gaining in complexity.
This presents us with unique opportunities, at this particular point in time, to re-introduce \textit{needs} into AI.
We believe that there is, fortunately, wind behind the sails of this effort.

\section{Drivers and Enablers}
Current successes (such as with deep learning and very large language models) will likely lead AI developers to continue down the paths they are on-- largely, adding more data and harnessing more powerful computing resources in order to improve results.
But there are other aspects and advancements that we also believe can/will push AI developers to look to \textit{needs} as a tool for building AI systems that are increasing useful and valuable to people. 

\textbf{1. The emerging sociotechnical imaginaries of \textit{needs-aware AI}}
\textit{Sociotechnical imaginaries}, the visions or values in related to a technology that are shared or common within the members of a community or a society, can influence how technologies are realized in practice (see e.g. \cite{jasanoff2013sociotechnical, jasanoff2009containing, felt2015five}). In this respect, \textit{sociotechnical imaginaries} are important non-human actors of \textit{co-production}. Generally, we contend, people are growing to expect more out of future AI than they expect today. From a \textit{human-centric} perspective, many individuals expect AI to \textit{satisfy needs}, as Kai-Fu Lee and Ben Shneiderman already suggest; \textit{needs} of different layers and levels (from individual \textit{needs} to collective, organizational, and societal \textit{needs}) as well as \textit{needs} associated with a sundry of physical, psychological, technical, and economic aspects of our lives.
These growing expectations can be framed as emerging sociotechnical imaginaries (within different communities and societies) regarding \textit{need-aware AI} systems.
We believe that these imaginaries can/will/are push/ing for new conversations and demands about the role of AI--and thereby \textit{needs}.

Integrating \textit{needs} into AI (i.e., \textit{for} \textit{AI}, \textit{through} AI, \textit{by} AI) can empower \cite{altEnduserEmpowermentDigital2020, humanEnduserEmpowermentInterdisciplinary2020} people in their relationship with AI systems of the future.
From developers to end-users, \textit{need-aware AI} could better partner with people to address \textit{needs}.
Likewise, with integrated \textit{needs} in AI systems, organizations (public and/or private) could in the future better prioritize and target resources based on formalized \textit{needs} rather than today's reliance on assumptions (e.g., those living in poverty must ``need''...) or ascriptions (e.g., you "need"...).
Equally, societies could benefit from the additional insights and guided actions of people and organizations with \textit{needs-aware} AI systems.
This is not a naive position, since AI co-produces \textit{needs} with people, we suggest that it is important for a diverse array of people (including \textit{needs} scholars, practitioners, and others) get out ahead of mainstream AI developments in terms of how \textit{needs} will be defined, prioritized, and measured.
This approach empowers humans, empowers organisations, empowers societies to set the \textit{needs} agenda for, through, and by AI.

Moreover, as discussed above, the \textit{AI technoscience}\footnote{The term \textit{technoscience} has a variety of meanings across different disciplines and communities. The meaning that we have in mind here is closer to the usage of this term in Science--Technology--Society (STS), e.g. \cite{latour1987science, de2011matters}.} (and the \textit{imaginaries} of \textit{AI as a technoscience}) can be very helpful in the study of \textit{needs}. This can be an important driver for the development of \textit{need-aware} AI as a scientific endeavour, besides all other individual, organizational, economic, and societal drivers. 

\textbf{2. Calls for Sustainable HALE AI}
In recent years, ethical concerns and debates regarding the development of AI systems have been frequent and intense. Besides the call to develop ethical AI systems (however they are defined), there are parallel interdisciplinary attempts regarding sustainability, human-centricity, accountability, and lawfulness of AI systems  (in industrial, academic, and socio-political levels). Privacy, fairness, trustworthiness, transparency, understandability, controllability, explainability, and many other aspects of AI systems have been widely discussed. Yet, 
when it is about practical means for implementing Sustainable Human-centric Accountable Lawful and Ethical (Sustainable HALE) AI systems, the communities have much less to offer in comparison to conceptual ideas or policy documents. 

From our cognitive systems to our values, from our responsibilities to our rights, \textit{needs} as one of the most fundamental aspects of \textit{human worlds} that play direct or indirect roles in shaping concepts and constructs that are important for the realization of Sustainable HALE sociotechnical systems. 
Therefore, we suggest that rethinking \textit{needs} by, for, and through AI can highly change our basic framings, assumptions, conceptualizations, and consequently, our policies, laws, roadmaps, guidelines, standards, frameworks, approaches and solutions.
This is in particular essential when (and if) we embody pluralist and inclusive positions that take all and every individual into account\footnote{see e.g. \cite{humanHumanCentricPerspectiveDigital2021} for a related cased-based discussion on privacy and marginalized people, which calls for a pluralist and inclusive approach towards development of technologies that consider all and every individual--as much as possible}. Every individual (person, organization, society, system, etc.) might \textit{satisfy their needs} differently. The ultimate integrated and well-functioning \textit{X-centricities} (e.g. joint human-centricity, ecology-centricity, society-centricity, etc.) and a higher level of \textit{non-discrete} sustainability, accountability, lawfulness, and ethicality can only be achieved if \textit{needs}, among others, are taken into account.

\textbf{3. AI won't [necessarily] let us wait}
The rapid advancement of AI and computing technologies over the last decade is putting pressure on researchers and practitioners in many other fields (from medicine and education to political science and zoology) to consider if they are prepared for the how AI will influence their work, and how their work may influence AI development.
From quantum computing to brain-computer interfaces, the technology is moving fast.
From digital humanities to social work, the capabilities of AI to inform and guide decisions is touching almost every field and discipline.

AI research is already actively applying the concept of \textit{needs} in their work, though almost always without recognition, definition, or clarification of what \textit{needs} are within their context.
Just as one example (out of many), \cite{qi2022conversational} suggests ``[i]n an ideal world, ConvAI [conversational AI] technology would help us build LUIs [language user interfaces] that allow users to convey their needs as easily as they would with other people.''
Yet, as this example illustrates, it is often assumed in AI research that, among other, (i) people can readily distinguish their \textit{needs} (i.e., what is necessary) from their wants or desires, (ii) people can easily (and in an understandable manner for AI) express their \textit{needs}, and (iii) that individuals' \textit{needs} should be treated as paramount in relation to those of others, or organizations/groups, or even societal \textit{needs}. 
Moreover, (iv) it is assumed that AI can easily assess what can \textit{satisfy} a specific \textit{individual}'s \textit{need(s)} in a specific \textit{context}.
Likewise, (v) how ConvAI should, in this example, act differently based on people's perceived \textit{needs} versus their other requests is an ignored aspect that requires further consideration by both the AI researchers and the communities that interact with the AI systems in the future.
Nevertheless, this lack of complexity in how \textit{needs} are considered and addressed in these early stages of AI research will, we suggest, set the precedent for how (or if) \textit{needs} are dealt with future AI. 
In other words, if \textit{needs} are not better defined and addressed soon, then assumptions and ascriptions about the \textit{needs} of others will dominate in AI development.

For those of us that study \textit{needs} (including philosophers, ethicists, educators, social workers, etc.) the recent developments in AI are creating pressure to move more quickly in their debates and deliberations, otherwise they might find that their efforts come too late to have influence in the future of AI and our society.
Market opportunities, scientific inquiry, and practical wants/desires--each are pushing some of us (i.e., \textit{needs} scholars and practitioners) to come to terms with what \textit{needs} are and how might AI serve human \textit{needs}.
As with raising a child, if you don't integrate ethics (including the integrated concept of \textit{needs}) at the beginning when they are relatively young, then it is much harder to add it into their knowledge-base and character later on.

This is not to suggest that a \textit{needs} community has to come to a single universal approach to \textit{needs} in AI, this is likely neither possible nor desirable.
But rather, interdisciplinary communities can/should soon create active resources and partner with AI developers; where AI developers can get guidance on how to bring the construct of \textit{needs} into their work and \textit{needs scholar and practitioners} can learn more about how AI can help advance our understanding of \textit{needs}.
In other words, it is expected that \textit{needs} scholar and practitioners become influential \textit{actors} in the co-construction of AI systems--before it is too late for these relationships to have substantial influence or impact. 

If, as we propose, \textit{needs} are essential to a future of AI that adds practical value to the lives of people, then a pragmatic approach to integrating \textit{needs} will most likely be found.
Even maybe, for instance, through brute-force by trying many, many different ways to estimate \textit{needs} until a workable approach is found.
This can, of course, be done with or without the ethical, philosophical, and humanistic qualities that are potentially available through broad interdisciplinary partnerships.
This timely push, however, we believe can serve as the impetus for interdisciplinary collaborations that put \textit{needs} into AI of the future.
Constructing a wide acceptance that \textit{re-thinking needs for, through, and by AI is essential for our societies} can be seen as a first important step. Clearly, answering the many \textit{how} questions ahead (and finding/constructing many more questions) is a matter of intensive joint interdisciplinary collaborations and co-creations, as next steps. 

\textbf{4. From Digitization to Digital Transformation}
Giving humans (i.e. users, customers, citizens, employees, etc.) a central role and considering their \textit{needs} and values while \textit{digital socio-techincal systems} (including AI systems) are co-produced is a fundamental distinguishing factors between \textit{digitization} (that refers more to the improvement of processes and efficiency) and \textit{digital transformation} (DX) that focuses more on \textit{humans' needs, values and experiences} \cite{ross2017don, humanACall2022}.
The last decades witnessed the impactful waves of \textit{digitization}. In the recent years, going more steps forward, governments, companies, organizations, and communities are more and more investing in \textit{digital transformation}. As a result, it is commonly accepted that digital transformation is going to fundamentally change our lives, relationships, perspectives, economies, political systems, science, and societies.

\textit{Needs}--in different levels and dimensions--are among the most important aspects of \textit{digital transformation} as a socio-technical program.
\textit{Needs} can inform our \textit{digital transformation strategies}, provide basis and assessment criteria for our \textit{digital transformation policies}, inform our \textit{digital transformation ethical frameworks}, and play a fundamental role in the real-world and applied co-creation of future sociotechnical systems. It is hard to imagine Sustainable Human-centric, Accountable, Lawful and Ethical digital transformation (Sustainable HALE DX \cite{humanHALEWHALEFramework2022}) without rethinking \textit{needs} into the development of digital sociotechnical systems.

Understanding and meeting diverse and contextual \textit{needs} and values through providing adaptive and personalized services \footnote{while considering other actors' and systems' \textit{needs} and values and taking sustainability concerns into account} are among the most common expectations of DX outcomes. AI is the most promising candidate technology to fulfil such expectations. Therefore, we argue, \textit{re-thinking needs into AI} can contribute towards better practices of digital transformation. Looking it from the other side, the increasing demand for DX, we suggest, is an important enable for \textit{re-thinking needs into AI}, since it makes the existing gap more than ever apparent.

Besides these, it is worth emphasizing that, among others, the significance and ethical requirements of \textit{needs} (both subjectively and objectively) in these increasingly impactful uses of AI are important to get right: among others,
``[b]ecause infrastructural technologies undergird systems of production, they begin changing societies and a people’s way of life by transforming the nature of work. The change occurs on two fronts: what people do for a living and how people do what they do''~\cite{barley2020work}.

\section{Threats and HALE Considerations}
While our call for \textit{re-thinking needs} \textit{for}, \textit{through}, and \textit{by} AI is partially motivated by the current and increasing concerns regarding sustainability, human-centricity, accountability, lawfulness, and ethicality of AI systems, we acknowledge and warn that--similar to many other approaches-- development of more and more \textit{needs}-aware AI system is not without potential negative consequences, if not done in a Sustainable HALE manner itself. Here, we reflect briefly on some of the most essential HALE considerations that should be taken into account when \textit{needs} (and \textit{needs satisfaction}) are re-thought \textit{for}, \textit{through}, and \textit{by} AI.

\textbf{1. AI Manipulation} 
As we discussed earlier, there is no doubt that technologies and humans have been co-producing each other throughout the history of homo sapiens. 
While our technologies have been very useful, and even might be used for knowledge generation, they did not have much \textit{knowledge} about us (if any)\footnote{Depending on our perspective on \textit{knowledge}, one might argue that some sort of \textit{knowledge} was \textit{embodied} in the past technologies.
However, here we are using the term \textit{knowledge} in a more common usage.}. However, digital technologies in general (and AI systems in particular) are increasingly able to construct different types of knowledge about the humans (and other actors or systems) that they are interacting with.
\textit{Needs}-aware AI might, in the future, possess vast \textit{knowledge} about what we \textit{need} and how we meet our \textit{needs}--more than any other technology.
And \textit{scientia potentia est} (i.e. \textit{knowledge} is \textit{power}), in particular if we consider the ubiquitous application of \textit{Ambient Intelligence}--\textit{everyware}.
As a result, a potential threat of \textit{needs}-aware AI systems, if not co-constructed and managed in a sustainable HALE manner, is their potential power to manipulate humans--and other systems from groups and organizations to societies--at their very core, i.e. their perceptions of their \textit{needs} and  \textit{need satisfaction}.

When referring to individual humans (or individual entities in general), it should be also considered that \textit{needs} are important for people (and entities), and in efforts to meet their \textit{needs} people (or entities) can find themselves in vulnerable positions. The threat of human manipulation based on their \textit{needs} (e.g., \textit{need X will be met but only if you do Y}, or \textit{your need is not `A' but rather it is really `B'}) is real. Today's social media companies already do similar (intentionally or unintentionally), with people routinely forfeiting some of their privacy in order to access content that meets their social ``wants'' (which many people perceive as ``needs''). As we increase  our understanding of how humans both identify and satisfy their \textit{needs} (especially using digital technologies), it will become increasing important for \textit{needs} to be integrated into policy, regulatory frameworks, and sociotechnical standards and guidelines that help protect \textit{human agency} and other rights. From \textit{needs-mining} to algorithms that price access to water, numerous new areas with the potential for manipulation are being created all the time.

\textbf{2. Our Imaginaries vs. Our Ignorances}
How we imagine \textit{needs-serving AI} is another area of potential threat to human agency. Just as movies like \textit{The Terminator} and \textit{The Matrix} have shaped public perceptions of AI, the imaginaries available to people for visualizing AI in the future are also important to what technologies eventually get developed by, for, and with AI.  What roles do we want for AI? What role should AI have in helping us identify and prioritize our \textit{needs}? What \textit{needs} do we want AI to help us satisfy? Should AI help us strike balances between individual, organizational, and societal \textit{needs}? Answers to these questions should also be part of the conversations that shape the next generation of ``imaginaries'' of AI.

Just as we don't want those making the laws or regulations also profiting from the policies they are creating, striking an appropriate equilibrium of \textit{needs}-aware AI (i.e., AI that informed by human \textit{needs} in design, implementation, and evaluation) and AI that is trying to satisfy specific human \textit{needs} is essential. 
Given that humans and AI now co-produce \textit{needs}, the boundaries of these relationships must be understood and guidance put in place to reduce the risk of human manipulation (based on \textit{needs}) or other threats. 
Additionally, satisfying one \textit{need} routinely equates to not satisfying some other \textit{needs}.
Making these determinations is about considering the context and priorities of the various \textit{needs} involved, and those trade-off are both challenging and potentially lucrative.
This is true both in prioritizing \textit{needs} and selecting ``satisfiers'' (i.e. ``which \textit{needs}'' through/by ``which satisfier'').

How is the latter discussion on the complexity of \textit{needs}-aware AI is related to the former reflection on \textit{imaginaries}? Here, we like to point to our \textit{ignorance} about the \textit{future} and \textit{complexity} of the world we are living in. While we advocate an active discussion and assessment of our \textit{imaginaries regarding \textit{needs}-aware AI}, we should also warn that the future world will not be exactly as we \textit{imagine} it. In  other words, \textit{socio-technical imaginaries} are important actors that contribute to the \textit{co-creation of needs-aware AI}, however, we should not be deceived by our imaginaries--forgetting our ignorance. That is why taking HALE considerations into account is so crucial. Even sometimes, it seems that calling for \textit{slower} co-construction of technologies that gives us more time to reflect, discuss, and manage our potential mistakes seems to be a wise position. It is hard to manage all--sometimes competing--actors involved in the development of AI, but it is not impossible to at least do our best to be actively and impactfully involved in the co-creation of the future AI. 

\textbf{3. Systems of Systems}
\textit{Needs} must be considered systemically (i.e. at the individual, group, society levels) since suboptimization is a continuous threat.
In other word, often satisfying a \textit{need} at one level (say, the individual level) can exacerbate \textit{needs} of either other individuals or of broader groups (such as organizations \textit{needs}, for instance, can also lead to exacerbating the \textit{needs} of some individuals. Therefore, all have to be considered as a holistic system.

The systems perspective is essential to reducing threats of bias, unfairness, and negative effects for marginalized people.
Adopting this perspective is not easy however, especially in relation to \textit{needs} where the relationships of \textit{needs} across and among these levels is both fundamental and always in flux.
Nevertheless, this is an important challenge that must be considered if AI is going to be capable of helping people meet their \textit{needs}.

\textbf{4. The HALE WHALE}
As advocators of \textit{Responsible Research and Innovation}, we are well aware that the list of potential considerations regarding \textit{Needs}-aware AI technologies can be very long. Here, just as a summary, we briefly provide a list based on HALE dimensions:
\begin{itemize}
    \item  From a \textit{H}uman-centric perspective, besides all complexities related to \textit{needs} assessment and satisfaction, ensuring \textit{transparency}, \textit{explainability}, \textit{understandability}, and \textit{controllability} of \textit{needs}-aware systems need special attention.
    
    \item From an \textit{A}ccountability perspective, it is important to include diverse human and non-human actors in the co-creation of \textit{needs}-aware systems, while \textit{responsibilities} and \textit{accountabilities} are well-defined. In other word, co-creation cannot be used as a means to distribute accountability in a way that no one hold responsibility any more.
    
    \item From a \textit{L}egal perspective, besides challenges of \textit{data protection}, \textit{consenting}, and \textit{consumer protection}, enforcing \textit{needs-awareness} as a prerequisite of advanced AI systems can be an important legal challenge. Moreover, it is clear that satisfying \textit{needs} should not invade legal rights and frameworks. Moreover, an important consideration is to set the legal (and practical) limits of \textit{needs}-aware systems. 
    
    \item from an \textit{E}thical perspective, besides all considerations that were discussed before--such as considerations regarding humans' \textit{agency}-- ensuring pluralism, inclusiveness and fairness of \textit{needs}-aware AI seems to be a challenging task.
    While the dominance of a specific set of approaches regarding \textit{needs satisfaction} does not seem to be an appropriate approach, a very critical consideration is how can we ensure that we do not assign ``equal validity''\footnote{i.e. \textit{everything goes!}} to every and all proposed approaches while supporting pluralism and inclusiveness.
\end{itemize}
Like a WHALE living in an ocean of different actors, a socio-technical digital system, such as a \textit{needs-aware} AI system, should function as a \textit{whole system} within and in-relation to other systems \cite{humanHALEWHALEFramework2022}. Keeping \textit{needs-aware} systems sustainable (from different perspectives) is a challenging consideration. Besides, all aspects that were mentioned above that can influence the sustainability of the \textit{needs}-aware systems, considerations and challenges regarding \textit{locality of knowledge}, \textit{federated computation}, \textit{distributed needs-aware systems}, and the application of \textit{synthetic data} expect to be addressed by interdisciplinary solutions, from philosophy and cognitive science to information systems and digital law (and beyond).  

\textbf{5. Stalling}
A final threat is the potential of not doing anything related to AI and \textit{needs} because we do not know how to do it. It is easy for challenges like those we are discussing here to overwhelm our capacity to plan and act, since none of this is easy.  But we posit that making a decision not to act (even if that is just starting conversations with colleagues about issues of \textit{needs} and AI) would be a regrettable choice. AI is continuing develop everyday, with literally more than a hundred new papers being share most days on arXiv.org alone. And if the professional communities across multiple disciplines don't come together, or wait to long to begin our conversations, then AI developers will answer many questions about the role of \textit{needs} in AI (many of which will be hard to change later).

\section{What Comes Next?}

In this article we have attempted to make the case for why we should integrate (conceptually,  computationally, and systemically) \textit{needs} \textit{for} AI, \textit{through} AI, and \textit{by} AI. There are, of course, other considerations, barriers, and enablers beyond our limited list here; nevertheless, our goal for making this case is as ``call to action''.
For \textit{needs} to be assimilated into the future of AI, we (i.e., \textit{needs} scholars and practitioners, AI researchers and developers, policymakers, and other actors) must begin our efforts to ensure that \textit{needs} are not left out of AI. In this vein, we propose the following: 

\textbf{1. Reconstruct the concept of ``need''}
How we choose to define \textit{needs} is a necessary step to move us toward the clearer conceptual and operational definitions of \textit{needs}, and the transparent and applied measurement\footnote{using quantitative, qualitative, and/or mixed methodologies} of \textit{needs} and \textit{needs satisfaction}.
We suggest that if AI developers are going to be able to utilize \textit{needs} in the [co-]design and implementation of AI systems (in order for those systems to help meet needs), then we must begin here as our foundation.

Today, however, definitions are not universally accepted across disciplines.
Many disciplines currently rely on definitions that attempt to set universal satisfiers of \textit{needs} (such as, \textit{self-actualization} or \textit{autonomy}) as the definition.
Others apply definitions rooted in deficiencies or gaps. 
Thus, the transformation of what \textit{needs} are, how we discuss \textit{needs} (both formally and informally), and how we create systems to measure \textit{needs} and \textit{needs satisfaction} (i.e. to \textit{`explicitize'} the knowledge of/about \textit{needs} and \textit{needs satisfaction}) will require systematic interdisciplinary efforts--to reach [working] agreements within and across many disciplines on common definitions that, e.g., can be applied from computer science to psychology, as easily as from social work to philosophy, and from law to economics.  

A common set of [working] definitions of what \textit{needs} are, we believe, is essential to creating AI systems that are \textit{needs-aware}.
In the end, these definitions may vary by context, for example, a health care AI using one variant and a criminal justice system using another.
Or through application, one or two operational definitions of \textit{needs} may be found as most useful and productive for AI systems.
In either case, a collaborative interdisciplinary approach may create a continuum from \textit{needs-blind} systems (i.e., those that disregard \textit{needs} in design and/or implementation) to \textit{needs-based} (i.e., those that prioritize needs in design and implementation as their most central aspect) -- with \textit{needs-aware} being a term to describe all different types of systems in the continuum that are not \textit{needs-blind}.

All of this is contingent, nevertheless, on having a definition(s) of \textit{needs} that can be communicated and applied. As well as a definition(s) that can applied for individual, organizational, and even societal level \textit{needs}.

\textbf{2. Create communities of action}
Progress on integrating \textit{needs} \textit{for}, \textit{through}, and \textit{by} AI in the future, depends on establishing broad interdisciplinary community(ies) that take actions.
Some of the required actions are traditional academic endeavours, such as writing articles, creating discussion forums, writing blogs, holding conferences (and conferences within conferences), teaching about \textit{needs} in courses, and of course conducting rigorous research. 
Others include supporting the development of AI literacy\footnote{which itself requires further research, see e.g. \cite{ngAILiteracyDefinition2021} for an initial attempt} in the social sciences and humanities so that the next generation of students, scholars, and practitioners are well versed in the technical and social aspects of conversations.
The interdisciplinary communities, however, can't remain isolated in academia; they must include public and private sector partners who can envision the positive roles of \textit{needs-aware} AI as well.  

\textbf{3. Co-construct the imaginaries}
One step, we propose, for invigorating professional dialogue (open to all disciplines) on the role of \textit{needs} in AI, is to introduce ``imaginaries'' (or scenarios) that illustrate the potential relationship of humans and AI systems in the future in terms of how they influence Kai-Fu Lee's vision of a future where ``AI will learn to serve human needs''.
These imaginaries offer a powerful and useful tool for disciplines to examine \textit{needs} within their particular context, as well as for interdisciplinary conversations to consider the appropriate role of \textit{needs} in AI systems that cross many boundaries.

\textbf{4. Promote federated computation and personal data protection}
In order to be sustainable in both individual and societal levels, we require access to data and knowledge in different layers. However, due to privacy and security concerns, this is not easily possible through traditional \textit{centralized} \textit{big data} or \textit{machine learning} approaches.
We propose that we can resolve this potential challenge by developing \textit{needs}-aware federated computation mechanisms, in which both privacy and agency of individuals are respected while data or knowledge is well distributed and communicated throughout the \textit{digital environment}. Many technical, legal, and societal challenges regarding implementation of such systems or \textit{eco-systems} should be resolved: an important mission for the future of our societies that cannot be achieved without collaborations between computer scientists, data scientists, philosophers, social scientists, lawyers, policymakers, corporations, NGOs, and many other actors involved.  

\textbf{5. Apply HCAI as an ethical AI framework}
If we use HCAI as a process for creating more ethical AI systems, then we can consider how/where/when \textit{needs} can be added into the HCAI processes to improve the results--both in terms of how the AI systems are designed, as well as how AI systems utilize \textit{needs} in their logic when making decisions/recommendations.  

HCAI can be done with a \textit{wants perspective} (e.g., what do people want), or from a \textit{[human]-needs perspective} (e.g., what are the priority \textit{needs}). 
At times these align (when people want satisfiers that will help meet their \textit{needs}), but routinely they do not align or are even in conflict (e.g., we want something that actually puts us in further \textit{needs}). 
Thus, the role of \textit{needs} in HCAI has to be considered carefully and researched in many contexts in order to identify the essential contributions of \textit{needs} in the process. 

Given the nature of \textit{needs}, we suggest that they are best integrated into HCAI through a mixed-methods lens.
The role of qualitative methods in the development of HCAI is already recognized (e.g., \cite{papakyriakopoulos2021qualitative}), and more specifically necessity and sufficiency are already commonly used measures in mixed-methods analysis approaches--including Qualitative Comparative Analysis \cite{ragin2014comparative, ragin2007fuzzy} and Necessary Conditions Analysis \cite{dul2020statistical}.
Formulas for calculating the probability of necessity and probability of sufficiency for given conditions have also been introduced by Pearl \cite{pearl1999probabilities}.
Through necessity and sufficiency considerations (or predictions) at multiple steps of the HCAI design process, we posit that rigorous approaches to \textit{needs} can be integrated \textit{for}, \textit{through}, and by \textit{AI}.

\textbf{6. Integrate \textit{needs} into ethical AI frameworks}
We propose that a critical step towards \textit{needs}-aware AI, is to introduce \textit{needs} into \textit{AI ethical frameworks}\footnote{as well as AI strategies}. Currently, multiple ethical frameworks are being proposed and debated (including, for example, those from the EU, USA, and China), none of which integrate the construct or measurement of \textit{needs} into their design.
The development of these initial frameworks lays for the foundation for future improvements, so missing the current opportunity to introduce \textit{needs} for, through, and by AI now will only make it more challenging to introduce it later--thus, this also has to be a priority for \textit{needs} and AI scholars and practitioners. 

\textbf{7. Rethinking \textit{Needs-awareness} to socio-economic perspectives and structures}
Lastly, our socio-economic perspectives and structures (from business models to power-structures, from public institutions to government policies) would benefit from  \textit{needs-awareness} [in AI and other sociotechnical systems]. For many, the introduction of \textit{needs} in their models, dialogues and decisions has been actively avoided since the complexity of \textit{needs} was beyond the capacity of the time, and for many others that might even  have conflict of interest with the realization of \textit{needs-aware} sociotechnical systems [including \textit{needs-aware} AI]. Nevertheless, emerging advanced technology enable us to now consider \textit{needs} for, through, and by AI.  

Along with all points mentioned above, we call for rethinking \textit{needs-awareness} to our socio-economic perspectives and structures in order to provide the broad structural supports required for the hard work of developing AI that serves \textit{needs}.

\section{Conclusions}
We have outlined above the major gaps, barriers, enablers and drivers for \textit{needs} (as a specific construct that can be described, measured and distinguished from other constructs) in the development of \textit{sustainable HALE AI}.
We have done so in hopes of igniting an interdisciplinary professional dialogue on the roles of \textit{needs}, and jump-starting real-world actions that can assist and guide the future of the AI that is capable of serving human \textit{needs}--a goal that can't be achieved without first coming to terms with our current lack of knowledge and understanding of our \textit{needs} or the \textit{needs} of others.
We hope that in response to this initial attempt to frame future conversations, others from philosophy, ethics, cognitive science (including psychology, neuroscience, cognitive biology, anthropology, etc.), political science, health, and many other disciplines that have been working on \textit{needs} and ways to assess \textit{needs} for decades, will share their perspectives in this dialogue.
At the same time, our desire is likewise to engage AI researcher and developers to engage with us and this topic, so that our efforts can lead to meaningful and impactful guidance and tools for creating future AI systems that meet our ideals and help us achieve our ideals. Ultimately, we hope that similar to \textit{rights} (e.g. human rights) that have become a fundamental aspect of our \textit{imaginaries} about technologies\footnote{i.e. it is hard to imagine anyone claiming to respect laws and ethics who explicitly advocates the violation of human rights through technology}, \textit{needs} also find their appropriate position in our shared visions: we imagine a world in which AI systems are co-created to satisfy [human] \textit{needs}; we imagine a world in which AI systems are--among others--planned, funded, designed, evaluated and judged based on the \textit{needs} they satisfy.
Please join the conversation by talking with your colleagues about \textit{needs}, integrating \textit{needs} into your work, and/or contributing editorials or articles about the roles for \textit{needs} in AI within your professional communities--and beyond. 

\bibliography{z-references} 

\end{document}